\documentclass[aps,prl,reprint]{revtex4-1}
\usepackage{graphicx}
\usepackage{dcolumn}
\usepackage{amssymb}
\usepackage{amsmath}
\usepackage{epstopdf}

\def\half{\frac{1}{2}}
\def\t#1{\textrm{#1}}

\begin{document}
\title{Dustball collapse and evaporation in standard coordinates}
\author{\"Ojvind Bernander}
\affiliation{ojvind.bernander@alumni.caltech.edu}
\date{February-September, 2021, edits 2022,2023}
\begin{abstract}

We consider evaporation alongside collapse for a dustball in standard (not comoving) coordinates.
A classical analysis gives the main result: an explicit metric (joined to the exterior) for the collapse.
The metric then provides the causal structure: light cones corresponding to the coordinate speed of light, $c$.
If the problem is perturbed from a pure dustball collapse, the solution can only be altered within the future light cones of such perturbations.

The metric tells us that $c_\textrm{interior}\ll c_\textrm{exterior}$.
Importantly, the speed at which the Schwarzschild radius shrinks during evaporation is intermediate between the two.
Thus a perturbation at the (shrinking) surface will only influence the exterior in finite time.
For example, if we assume evaporation to be a process located at the dustball's surface, we can model it as a series of perturbations to the classical solution.
In this modified solution, the interior is not altered as the evaporation process eats into a frozen interior from the outside in;
the singularity and a region near it are no longer present;
infalling particles don't cross the (shrinking) dustball boundary;
the metric coefficients don't change sign with time,
and therefore timelike world lines can exist at constant radius inside the (shrinking) Schwarzschild radius and emerge at late times (no absolute horizon forms).

Research on evaporation alongside collapse typically study collapsing shells and apply quantum field theory.
This paper differs in that it studies an entire dustball and uses classical general relativity to constrain evaporation models.
\end{abstract}

\maketitle

\section{Introduction}

\begin{figure}[b]
\includegraphics[width=8.7cm,keepaspectratio]{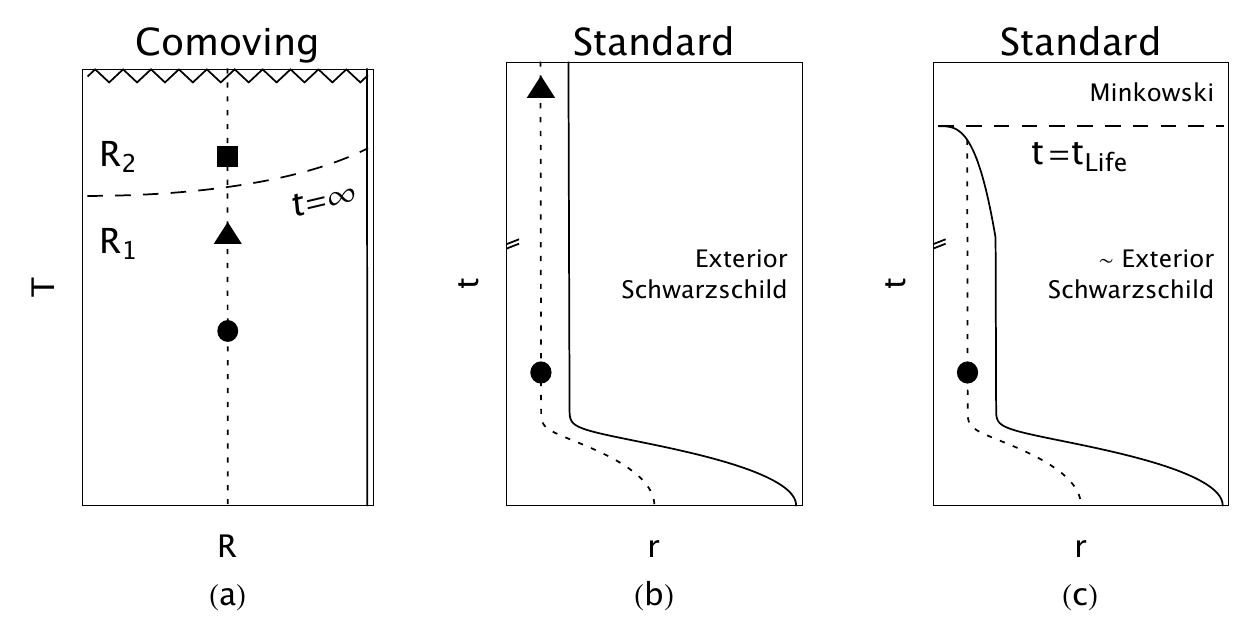}
\caption{\label{fig:1}
Collapsing dustball.
Solid lines: dustball surface.
Dotted line: trajectory of an interior dust particle with three events marked (circle, triangle, square).
(a) Classical collapse in comoving coordinates $(T,R)$. Jagged line: singularity. Dashed line divides regions $R_1$ and $R_2$; only $R_1$ is mapped to standard coordinates.
(b) Classical collapse in standard coordinates $(t,r)$, failing to cover region $R_2$.
(c) Evaporating collapse in standard coordinates. Evaporation is complete at time $t=t_\t{Life}$.
Upper portion of $t$ axis is compressed in (b) and (c).
}
\end{figure}

Oppenheimer and Snyder analyzed the collapse of a spherically symmetric ball of dust.\cite{Oppenheimer39}
In their {\em comoving coordinates,} dust particles are at rest, with $T$ being the particle's proper time,
and $R$ its (unchanging) radial location. Their metric is

\begin{equation}
d\tau^2 = dT^2-Q^2(T)(\frac{1}{1-kR^2}dR^2 + R^2d\Omega^2),
\label{eq:metric:comoving}
\end{equation}

\noindent
where $d\Omega^2=d\theta^2+\sin^2\theta\,d\phi^2$.
Figure~\ref{fig:1}(a) shows {\em classical} (non-evaporating) collapse.
The dustball's surface is to the right (vertical solid line),
a sample particle is in the center (dotted line with three events indicated),
ending in a singularity
(squiggly line at top)
where $Q(T_\t{max})=0$.

For a spherically symmetric spacetime, we can always write the metric in {\em standard coordinates} $(t,r,\Omega)$,
where the metric has the form:
\begin{equation}
d\tau^2 = A(r,t)dt^2-B(r,t)dr^2 - r^2d\Omega^2,
\label{eq:1}
\end{equation}

\noindent
which is often more intuitive and resemble both the Minkowski metric in spherical coordinates, and the the Schwarzschild metric.
(Exterior to the dustball, the metric equals the Schwarzschild metric.)
We will present $A$ and $B$ for classical collapse.

These standard coordinates have a drawback for classical collapse:
they don't cover the entire manifold, but only the region $R_1$ in Figure~\ref{fig:1}(a).
As seen in Figure~\ref{fig:1}(b), the third event (square symbol in region $R_2$) is absent,
and we are left with a ``frozen star."

At first blush it may seem that standard coordinates are also deficient when analyzing {\em evaporating} black holes:
collapse is fast in comoving time $T$, while evaporation is slow in standard time $t$;
therefore, collapse may seem to complete {\em before} evaporation has much effect.
However, $T$ and $t$ must not be compared in this way, as they appear in very different coordinate systems.
More importantly, events in region $R_2$ are space-like separated from the (post-collapse) exterior and much of the rest of the dustball,
so the word {\em before} would be ambiguous.
Rather than collapse-{\em and-then}-evaporation, we must assume collapse-{\em alongside}-evaporation, unless proven otherwise.\cite{Frolov97}
This has been done previously for collapsing shells
\cite{Boulware-1976,Gerlach-1976,Alberghi-2001,Vachaspati-2007,Barcelo-2008,Dai-2016,Baccetti-2017,Paranjape-2009,Mann-2022};
here we do it for a dustball.

Our opinion is that standard coordinates better visualize the processes of collapse and evaporation.
In particular they may reduce the risk of mistakenly including regions of the classical-hole solution that are absent in the presence of evaporation.
This is suggested in Figure~\ref{fig:1}(c):
We know that at times $t>t_\t{Life}$ spacetime is empty and flat, and so we suspect that the second event (triangle) ought to be absent,
along with all of region $R_2$, as we argue in a later section.

Also note these important points about standard coordinates, which we obtain by setting various differentials to zero in equation~\eqref{eq:1}:
\begin{enumerate}
\item
$2\pi r$ measures circumference of static equatorial rings at $r$ ($dt=dr=d\theta=0,\,\theta=\pi/2$).
We can imagine physical rings of negligible mass that remain at constant $r$ as long as $A,B>0$.
(If $A,B<0$, as is the case for the {\em interior} Schwarzschild metric, then it is not possible for matter to remain at constant $r$:
timelike world lines then require $r$ to decrease with $t$.)
\item
For radial null geodesics ($d\tau=d\Omega=0$), we can define a coordinate speed of light, $c_r=\sqrt{A/B}$),
limiting the coordinate speed of particles.
\item
Clocks at fixed locations
($dr=d\Omega=0$)
measure time $d\tau=\sqrt{A}\,dt$.
\end{enumerate}

The first point allows us to place observers on rings of constant $r<r_S$, where $r_S$ is the Schwarzschild radius,
and they will remain in place for all finite $t$.

The second point puts a limit on how deeply the effects of a process at the dustball's surface can penetrate in finite $t$.
We shall see that $c_r$ approaches zero exponentially, and penetration essentially vanishes.
In contrast, $|\dot{r}_S|$, the speed at which the Schwarzschild radius shrinks by evaporation, is vastly greater than the interior $c_r$.
Thus, we will see that an evaporation model located at the surface has the shrinking surface erode the frozen interior from the outside in,
and region $R_2$ is not realized.
This model may help build intuition for other research, for example on the information paradox and firewalls.

This paper has two main sets of results.
The first is for classical dustball collapse in standard coordinates, for early and late times.
The second regards limits on how an evaporation perturbation can propagate, suggesting that neither a horizon, nor a singularity forms.

\section{Background}

Our notation is modified from that of section 11.9 in Weinberg,\cite{Weinberg72} who recapitulates the dustball collapse of Oppenheimer and Snyder.

{\bf Dustball.}
The mass is $m$, as it would appear in the exterior vacuum in the Schwarzschild metric.
Initially, the dust is at rest, the proper radius is $r_0$, and the proper density is $\rho_0$ throughout.
(By ``proper radius" we mean the value $C/2\pi$ where observers on a sphere of constant radius measure the circumference $C$ of a great circle.)

For later convenience, we introduce $n=r_0/r_S$, where $r_S=2m$ is the Schwarzschild radius of a mass $m$.
We then have $\rho_0 = (3/4\pi)(m/r_0^3) = (3/32\pi)(1/n^3m^2).$

\begin{figure}[b]
\includegraphics[width=8.7cm,keepaspectratio]{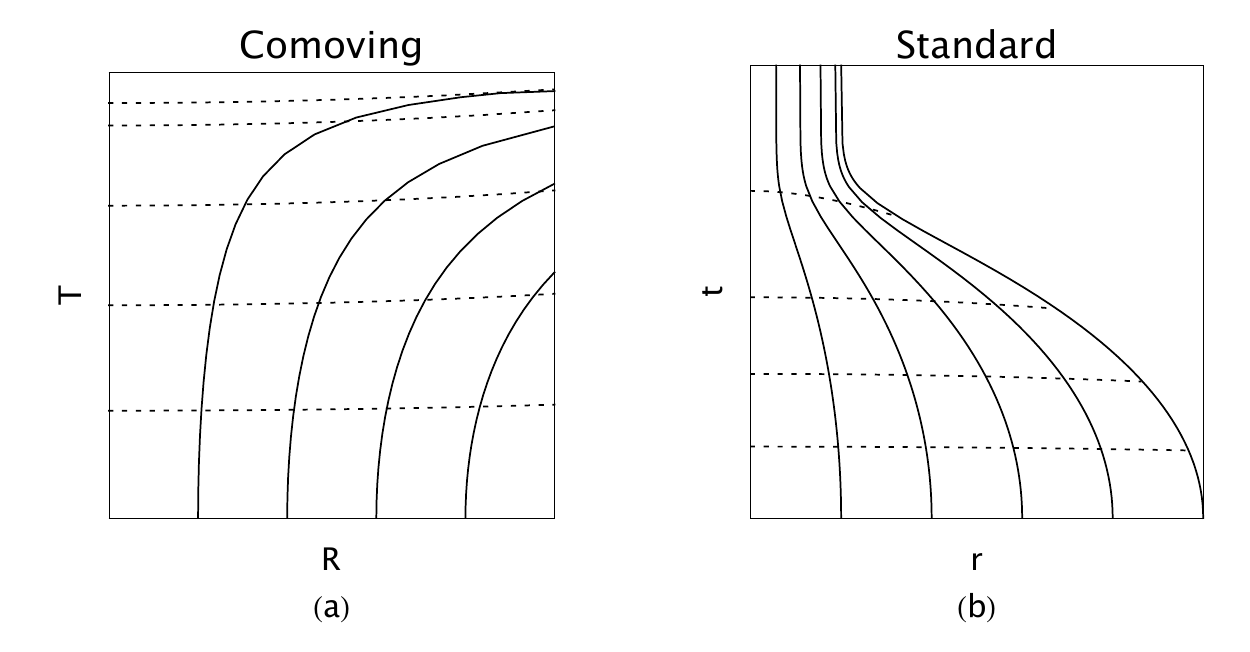}
\caption{\label{fig:2}
Coordinate curves.
(a) In the $R$--$T$ plane, curves of constant $r$ (solid) and $t$ (dotted).
(b) In the $r$--$t$ plane, curves of constant $R$ (solid) and $T$ (dotted).
Constant values are evenly spaced.
}
\end{figure}

{\bf Comoving coordinates, }$(T,R,\Omega)$.
In these coordinates, dust particles are always at rest, as indicated by vertical lines in Figure~\ref{fig:1}(a)
(interior particle dotted, surface particle solid).
The metric is given by equation~\eqref{eq:metric:comoving}, where the space part is scaled by $Q(T)$.
$Q$ and $T$ are related parametrically by the cycloid equations:

\begin{eqnarray*}
  & Q(\psi)=\half(1+\cos\psi)
  \\
  & T(\psi)=n^{3/2}m(\psi+\sin\psi)
\end{eqnarray*}

\noindent
From start to end, $\psi$ evolves from $0$ to $\pi, ~ Q$ from $1$ to $0$, and $T$ from $0$ to $T_\t{max}=\pi\,n^{3/2}m$.
$R$ ranges from $0$ to $r_0$ at all times.
Proper radius is $Q(T)R$ with a singularity for $Q(T_\t{max})=0$.
Proper density evolves as $\rho(t)=\rho_0/Q^3(T)$.

{\bf Standard coordinates, }$(t,r,\Omega)$.
Weinberg provides us with a transform

\begin{eqnarray}
   & r=Q(T)R
   \label{eq:rR}
\\ & t(S) = r_0\sqrt{n-1}\int_S^1\frac{nx}{nx-1}\sqrt{\frac{x}{1-x}}dx
   \label{eq:tS}
\\ & S(T,R) = 1 - \sqrt{\frac{n-f(R)^2}{n-1}} (1-Q(T))
   \label{eq:STR}
\end{eqnarray}

\noindent
where S is an intermediate variable, and we solve for $t(S)$ in closed form below.
$f$ is a rescaling of either radial variable: $f = R/r_0 = R/(2nm) = r/(r_0Q)$.
In Figure~\ref{fig:2} we see how the coordinates relate inside the dustball.

Weinberg also provides the resulting metric

\begin{eqnarray}\label{eq:weinberg-metric}
   & A(Q,f,S) = \frac{Q^2(nS-1)^2}{S^3(nQ-f^2)n}\sqrt{\frac{n-f^2}{n-1}}
   \label{eq:A}
\\ & B(Q,f) = \frac{nQ}{nQ-f^2}
   \label{eq:B}
\end{eqnarray}

\noindent
Joining this to the Schwarzschild metric at the dustball surface gives
$\rho_0 = (3/4\pi)(m/r_0^3)$, which we previously stated without motivation.

Weinberg stops at this point.
Note that the equations~\eqref{eq:rR}-\eqref{eq:STR} are coupled;
also, other than $t(S)$ they are not in closed form.
$A$ and $B$ are explicitly functions of $Q,f$ and $S$ rather than directly of $t$ and $r$.
To quote Weinberg, this is a mess.
We aim to simplify the mess.

\section{Methods}

We derive some expressions in closed form, such as $T(Q)$ and $t(S)$, while others cannot be expressed in closed form, such as $Q(T)$ and $S(t)$.
We use the software program Mathematica to numerically evaluate both types, and to monitor precision and increase it when needed.
Mathematica is also used as an aid to solve equations and integrals and general algebraic manipulation,
though manual simplification using trigonometric and hyperbolic identities resulted in more compact expressions.
To select among multiple roots (for example when solving common quartic polynomials),
numerical evaluation of known boundary points always revealed a unique choice.

To derive closed-form approximations for large $t$ we inspected expressions for dominant terms (see the appendix) and checked against very high precision numerical evaluations.

We use geometrized units such that $G=c=1$;
time is then measured in units of $1/c_\t{SI}$ seconds and mass in $c_\t{SI}^2/G_\t{SI}$ kg;
Planck's constant takes on the numeric value $\hbar=2.6122\,10^{-70}$.

\section{Results: Classical collapse}

\begin{figure}[b]
\includegraphics[width=8.7cm,keepaspectratio]{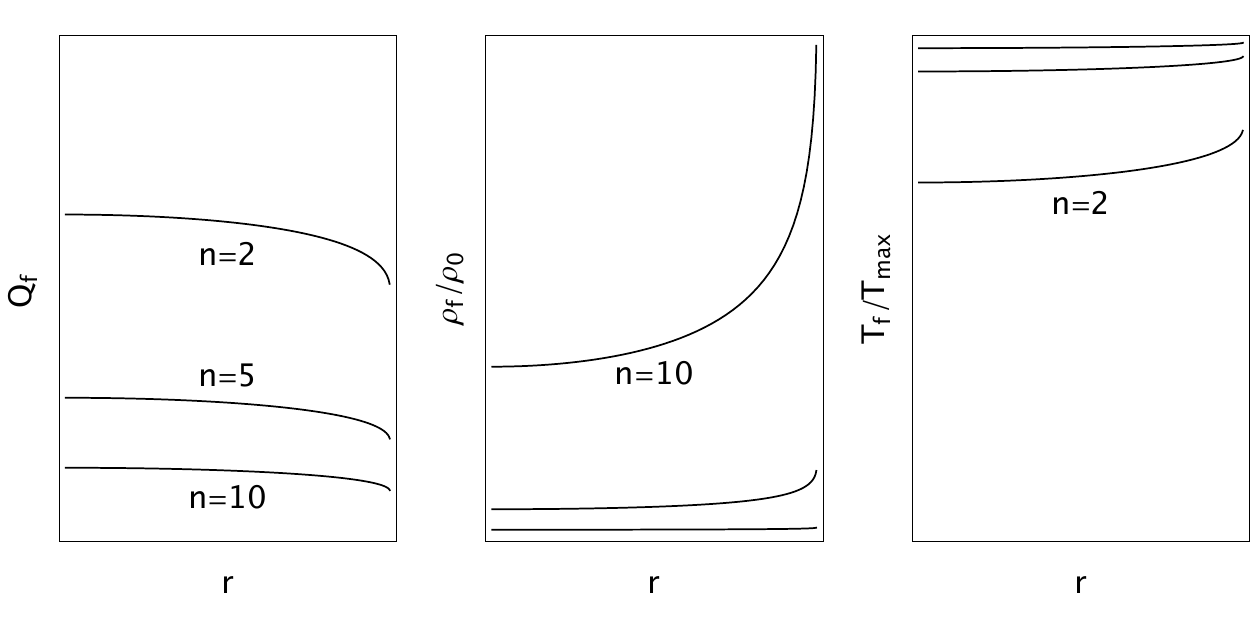}
\caption{\label{fig:3}
Final ``frozen" values as $t\rightarrow\infty$
for classical collapse. $n=2,5,10$.
}
\end{figure}

Many results are presented in this section and are then used in the following section on evaporating collapse.
Results that are particularly useful include:
all quantities remain finite and non-zero for finite $t$;
$A$ and $B$ remain positive for all finite $t$;
$c_r$ decays exponentially toward zero;
the dustball's surface decays exponentially toward the Schwarzschild radius $r_S$.

{\bf Auxiliary expressions.}
In order to express $A$ and $B$ in standard coordinates, we must first find $S(t),~Q(t,r)$ and $f(t,r)$,
to plug into equations~\eqref{eq:A}-\eqref{eq:B}.

Equation~\eqref{eq:tS} is solved and simplified to

\begin{eqnarray}
   \label{eq:tS2}
   & t(S) = 4m\,\t{arctanh}(\sqrt{\frac{1-S}{(n-1)S}}))
\\ & +2m\sqrt{n-1}[ (2+n)\arccos(\sqrt{S}) + nS(1-S) ]
   \nonumber
\end{eqnarray}

\noindent
which is monotonic (see Appendix).
The inverse $S(t)$ must be computed numerically.
Inspecting the integral, we note that $S(0)=1$ and $S(\infty)=1/n$.

To obtain $Q(t,r)$ we substitute $f=r/(r_0Q)$ into equation~\eqref{eq:STR} and obtain a quartic polynomial in Q.
The appendix gives the general solution, somewhat compactly as a nested chain of functions of the polynomial's coefficients
(which in turn depend on $n,m,S(t)$ and $r$).
We select the correct root by entering trial values and comparing to $Q(T)$ in the cycloid equations.

We now have, trivially,

\begin{equation*}
f(t,r) = \frac{r}{r_0Q(t,r)}
\end{equation*}

{\bf Frozen state.}
For $t=\infty$, we then have

\begin{eqnarray*}
  & Q_f(r)=Q(\infty,r)
  \\
  & \rho_f(r)=\rho_0/Q_f^3
  \\
  & T_f(r) = n^{3/2}m(\psi+\sin\psi),~\psi=\arccos(2Q_f-1)
\end{eqnarray*}

\noindent
(The subscript $f$ stands for {\em final} or {\em frozen}.)
These are graphed in Figure~\ref{fig:3} for various $n$.
Note that $Q$ has evolved less at small $r$, as might be expected:
clocks tick more slowly deep inside a sphere than at its surface.
Consequently, both the density $\rho_f$ and proper time $T_f$ are smaller at small $r$.

$T_f(R)$ divides spacetime into regions $R_1$ and $R_2$ in Figure~\ref{fig:1}(a).
For classical collapse, this division is of no special interest, as noted earlier.
Nor does the frozen dynamics in standard coordinates imply that dynamics cease in other coordinates:
for example, in comoving coordinates evolution continues into region $R_2$.

{\bf Worldline of dustball boundary, $r_b(t)$.}
We join the interior metric to the Schwarzschild metric at the dustball boundary, $r_b$,
and so we need an expression to find its radial location at different times $t$.
The inverse is

\begin{eqnarray}
   & t(r_b) = 4m\,\t{arctanh}\sqrt{\frac{2mn-r_b}{(n-1)r_b}}
   \nonumber
\\ & +2m\sqrt{n-1}(2+n)\arctan\sqrt{\frac{2mn}{r_b}-1}
   \label{eq:trb}
\\ & +\sqrt{(2mn-r_b)(n-1)r_b}
   \nonumber
\end{eqnarray}

\noindent
(This exact expression can be obtained by the integral $\int_{r_0}^{r_b}v^{-1}(r)\,dr$,
where $v(r)$ is the speed of an object falling freely from $r_0$;
see for example chapter~3 in \cite{taylor00}.)
$r_b(t)$ is evaluated numerically as the inverse of $t(r_b)$.

\begin{figure}[t]
\includegraphics[width=8.7cm,keepaspectratio]{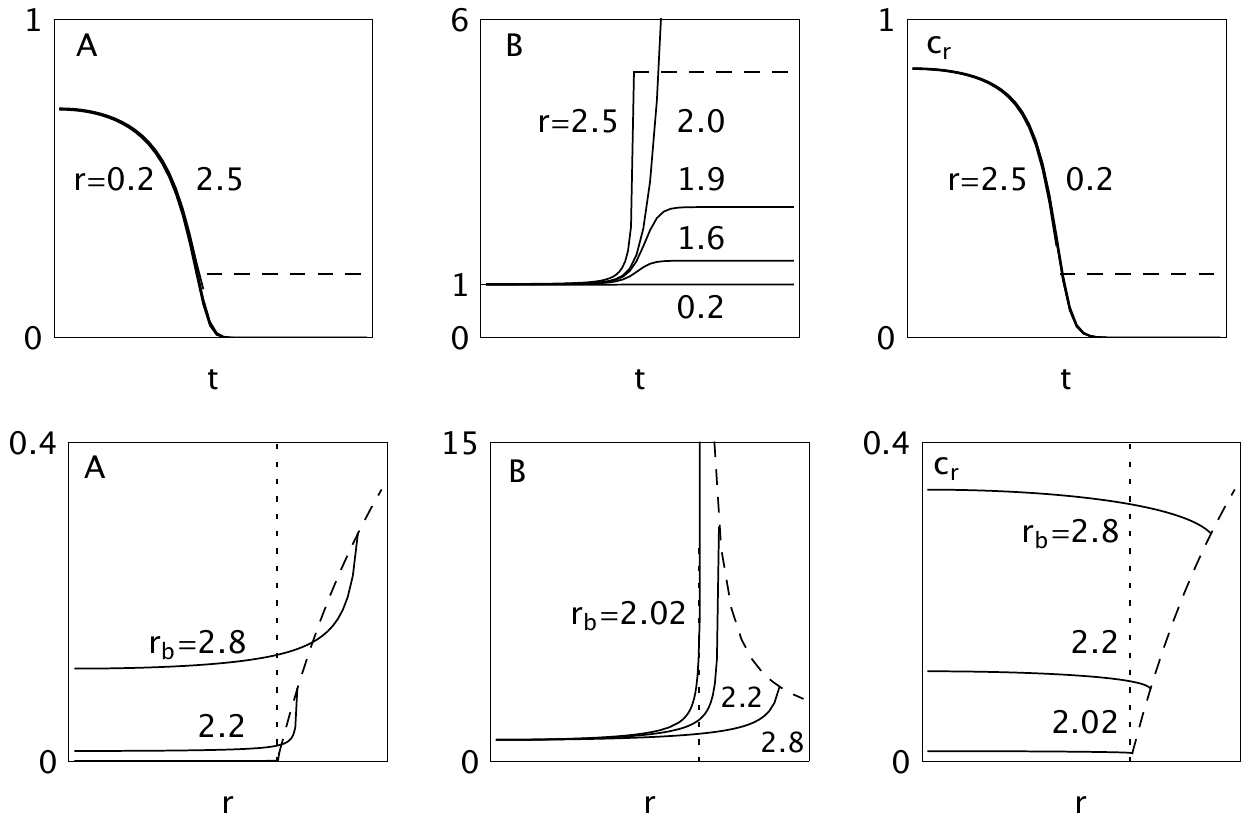}
\caption{\label{fig:4}
$A,B$ and $c_r$ as a function of $t$ at fixed $r$, top row,
and as a function of $r$ at fixed time (labeled by the location $r_b$ of the dustball surface), bottom row.
Schwarzschild radius $r_S$ dotted.
External Schwarschild metric dashed.
$m=1,~n=5$.
}
\end{figure}

{\bf Metric and $c_r$ for small $t$.}
From equations~\eqref{eq:A}, \eqref{eq:B}:

\begin{equation}
c_r(Q,f,S) \equiv \sqrt{A/B} = \frac{nS-1}{nS}\sqrt{\frac{Q}{S}}(\frac{n-f^2}{n-1})^{1/4}
\label{eq:cr}
\end{equation}

\noindent
We numerically evaluate $A,B$ and $c_r$ as functions of $(t,r)$ to high precision.
Figure~\ref{fig:4} shows results for a dustball with mass $m=1$ and $n=5$, and so $r_S=2$ and the initial $r_0=10$.

In the top row, we choose some values inside ($r<2$), outside ($r=2.5$) and exactly at $r=r_S=2$.
$A$ decreases exponentially toward $0$  with a time constant $\tau_A=m$, which agrees with the approximate expression below.
$B$ increases to plateaus inside the dustball;
the only exception is exactly at $r=r_S$ where $B(r_S)$ increases without limit;
at $r=2.5$, the dustball surface eventually moves past, and we are left with a plateau at the Schwarzschild value (dashed).
$c_r$, like $A$ approaches zero exponentially, but with a time constant $\tau_c=2m$.
In the exterior, for $r=2.5, ~A$ and $c_r$ have non-zero values.

In the bottom row we choose three times, at which $r_b$ takes on the values $2.8,~2.2$ and $2.02$.
The Schwarzschild radius is dotted, and the exterior Schwarzschild metric is dashed.
Both $A$ and $B$ increase sharply just inside the surface, but their ratio is more level, as reflected in $c_r$.
As noted in the introduction, proper time evolves as $\sqrt{A}\,t$;
this evolution is slower for smaller $r$, consistent with $Q$ being larger at smaller $r$.

Note that all three quantities remain positive.

Also note that all three quantities are continuous at the surface $r_b$, outside which the metric becomes the Schwarzschild metric.
Derivatives, however, are discontinuous at $r_b$, reflecting the discontinuity in dust density.

{\bf Approximations for large $t$.}
We want approximate expressions when $t\rightarrow\infty$.
Here we present the results;
derivations are outlined in the Appendix;
as sanity checks we numerically compared them to the expressions above.

The following quantities exponentially approach an asymptote:

$$
\begin{array}{lll}
   S(t)     & \approx  k_0e^{-t/2m} & +\, 1/n
\\ r_b(t)   & \approx  k_1e^{-t/2m} & +\, 2m
\\ Q(t,r)   & \approx  k_2e^{-t/2m} & +\, Q_f(r)
\\ A(t,r)   & \approx  k_3e^{-t/m}
\\ B(t,r)   & \approx  k_4e^{-t/2m} & +\, B_f(r)
\\ c_r(t,r) & \approx  k_5e^{-t/2m}
\end{array}
$$

\noindent
where constants depend on $r$ and parameters $n,m$:

$$
\begin{array}{lll}
   k_0 & = r(n-1)n^{-2} e^{ \sqrt{n-1}(\sqrt{n-1}+(2+n)\,\t{arcsec}\,(\sqrt{n}) }
\\ k_1 & = 8(n-1)m\,n^{-1}e^{ \sqrt{n-1}(\sqrt{n-1}+(n+2)\arctan\sqrt{n-1}) }
\\ k_3 & = (k_0^2n^4Q_f^2)(nQ_f-\beta r^2)^{-1}\sqrt{ (n-\beta r^2)/(n-1) }
\\ k_4 & = -3k_0k_2nr^2\beta(\beta r^2-nQ_f)^{-2}
\\ k_5 & = \sqrt{nQ_f} ((n-\beta r^2)/(n-1))^{1/4}
\\ \beta & =(2nmQ_f)^{-2}
\end{array}
$$

\noindent
$k_2$ has a complicated expression, arising from the fact that $Q$ is the solution of a quartic polynomial.
The full, nested expression is given in the appendix.

Equation~\eqref{eq:B} with $Q=Q_f$ gives $B_f$.
This expression lacks an asymptote at $r=r_S$,
and so the approximations for $B$ and $c_r$ are only valid for $r<r_S$,
while the approximations for $A$ and $Q$ are valid for all $r<r_b$.

Again note that $A,B$ and $c_r$ remain positive for all finite $t$.
The asymptotes for $A$ and $c_r$ are zero.
All time constants are $2m$, except for $A$.

\section{Results: Evaporating collapse}

We consider numerical values for a dustball with $m=3000, ~n=3$, approximately the mass and size of a large neutron star.

When a black hole evaporates by Hawking radiation~\cite{Hawking-1975} we use the following standard formulas:

\begin{eqnarray*}
   & m(t) = (m_0^3-3\alpha t)^{1/3}
   \nonumber
\\ & \dot{m} = -\alpha m = -\alpha(m_0^3-3\alpha t)^{-2/3}
   \nonumber
\\ & t_\t{life}=m_0^3/(3\alpha)
   \nonumber
\\ & \alpha=\hbar c^4/(15360\pi G^2).
   \nonumber
\end{eqnarray*}

\noindent
The Schwarzschild radius is now a function of time: $r_S(t)=2m(t)$.
Initially, the speed at which its Schwarzschild radius shrinks is

\begin{equation}
  |\dot{r}_S(0)| = 2|\dot{m}(0)| =  3.6\times10^{-73} \, m/s
\end{equation}

\noindent
after which the speed slowly increases.

\begin{figure}[t]
\includegraphics[width=8.7cm,keepaspectratio]{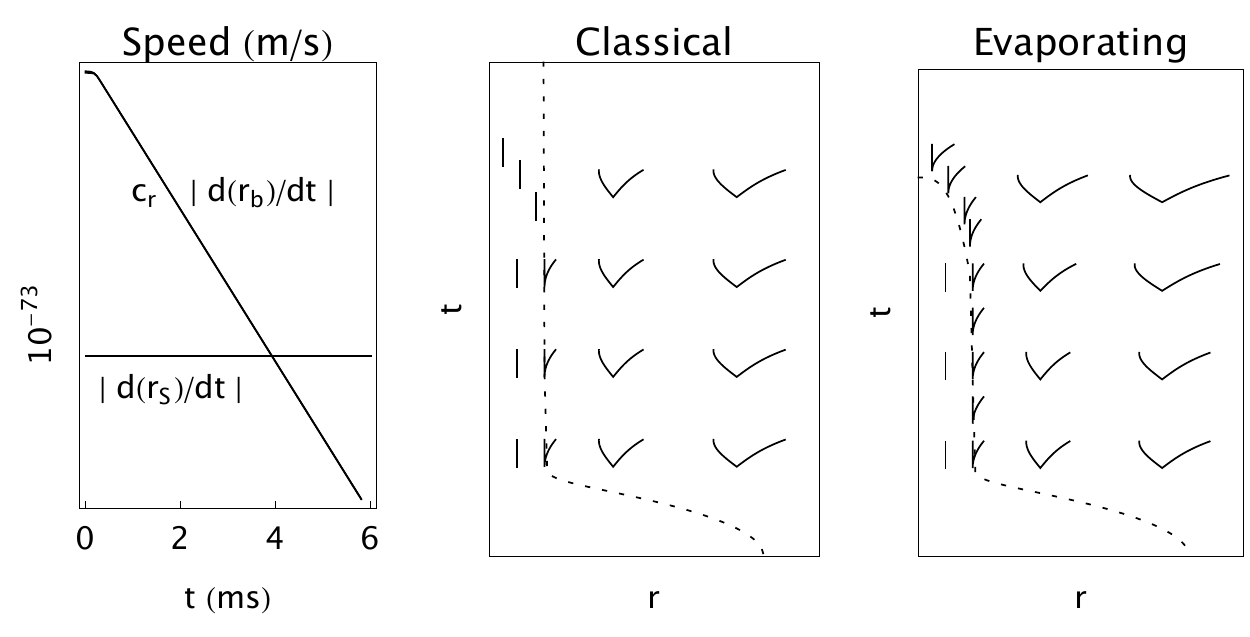}
\caption{\label{fig:5}
Left: three speeds compared (logarithmic y axis).
A dustball with $m=3000,~n=3$ begins classical collapse at t=0.
Speed of light at $r_S$ and the speed of the dustball surface, $d(r_b)/dt$, both decay exponentially with a time constant of $2m$.
The evaporation rate, $d(r_S)/dt$ is initially almost constant, increasing at very large $t$ (not shown).
Middle and right:
null geodesics (solid) determine the region of influence (ROI) of a perturbation,
shown here schematically as bent light cones.
Near the surface, ROIs are tilted such that they essentially contain only points exterior to the perturbation.
}
\end{figure}

\begin{figure}[b]
\includegraphics[width=8.7cm,keepaspectratio]{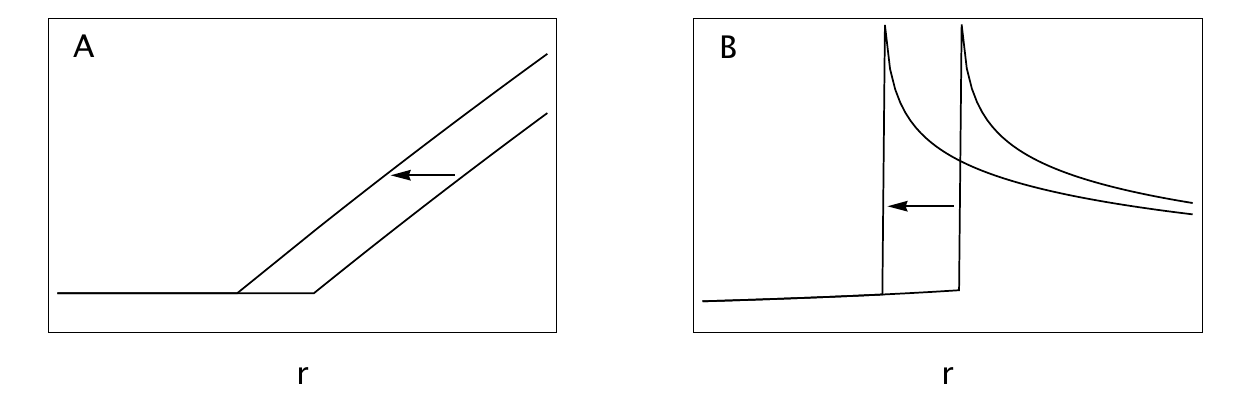}
\caption{\label{fig:6}
Metric coefficients $A$ and $B$  at two times,
in a model where the process that causes evaporation occurs in a narrow shell (of order $\Delta r$) at the dustball surface.
}
\end{figure}

{\bf Speed of shrinkage greatly exceeds interior $c_r$.}
We compare this speed with two other speeds in Figure~\ref{fig:5} (left panel):
the radial speed of light, $c_r$, just inside the surface, and the speed at which the surface $r_b$ approaches $r_S$.
$\dot{r}_b$ almost exactly overlaps with $c_r$ as is to be expected:
near the Schwarzschild radius, infalling dust approaches the speed of light.

It is noteworthy that after only 4~ms the largest speed is that of the shrinking Schwarzschild radius, $|\dot{r}_S|$.

This suggests that the classical expression for $r_b(t)$ stops being valid after about $4~ms$.
Similarly, the classical expressions for $A,B$ and $c_r$ may not be valid for $r>r_S(t)$.

{\bf Skin depth.}
A particle with velocity $v_0e^{-t/\tau}$ travels a maximum distance of $v_0\tau$ in finite $t$.
Thus a perturbation to the (internal) metric at time $t$ propagates at most a distance $d_\t{max}=2mk_5e^{-t/2m}$,
which after a few milliseconds is effectively zero.

{\bf Exterior }$c_r~${\bf greatly exceeds speed of shrinkage.}
For the Schwarzschild metric, we saw in the last panel of Figure~\ref{fig:4}
that the radial coordinate speed of light, $c_r$, increases steeply with distance outside the surface.
At $\Delta r\approx5\times10^{-53}\,m$ outside $r_S$,
$c_r$ equals $\dot{r}_b$.

{\bf Perturbations and their regions of influence.}
Null geodesics define regions of influence (ROIs).
A point perturbation, such as a firecracker, can modify the solution only within its future ROI.
The middle panel of Figure~\ref{fig:5} shows schematic ROIs as bent light cones.
Outside $r_S$, outgoing rays speed up and ingoing rays slow down and approach a vertical asymptote at $r_S$,
tilting the ROIs.
Inside $r_S$, the small value of $d_\t{max}$ has the ROIs collapse to (almost) vertical lines.

Just outside $r_S$, ingoing rays again stay put as (almost) vertical lines,
while outgoing rays can escape.
The slope of the inner null geodesic is steeper than the slope of $r_S$ for an evaporating black hole
(per the asymmetry of speeds discussed above).

This suggests that a perturbation at the surface can propagate outward, but not inward.

{\bf Modeling of evaporation.}
Evaporation is a non-classical process, often thought of as occurring at or near $r_S$.
Let us start with the classical solution and model evaporation as a perturbation at $r_S$.
The right panel of Figure~\ref{fig:5} shows nine such evaporation events along the shrinking $r_S$.
Their ROIs open outward, modifying the exterior part of the solution only.

This suggests that $A$ and $B$ would evolve qualitatively as in Figure~\ref{fig:6}.
An edge between a frozen interior and a near-empty exterior moves toward smaller $r$.
For the exterior, $\dot{B}<0$, and so the $t$--$r$ component of energy-momentum, $T^{01}=-\dot{B}/(AB^2r)$ is positive,
which is consistent with outgoing radiation.
The exterior is approximately Schwarzschild with a slowly decreasing mass $m(t)$;
it could be modeled with Vaidya null dust\cite{Vaidya51}, though Hawking radiation is not necessarily radial,
and its atmosphere may contribute pressure to the energy-momentum tensor.

From the above, we only expect the edge itself to move inward, but not the resulting disturbance to the classical solution,
because the edge itself would move faster than the internal coordinate speed of light.
Externally, in contrast, at a distance greater than $\Delta r, ~ c_r$ exceeds $|\dot{r}_S|$,
and so the disturbance of the evaporation process can propagate away from the black hole.
This leaves a narrow band (shell) at the moving edge with a width of some multiple of $\Delta r$ that could be modeled in detail;
this paper does not offer such a detailed model.

\begin{figure}[b]
\includegraphics[width=8.7cm,keepaspectratio]{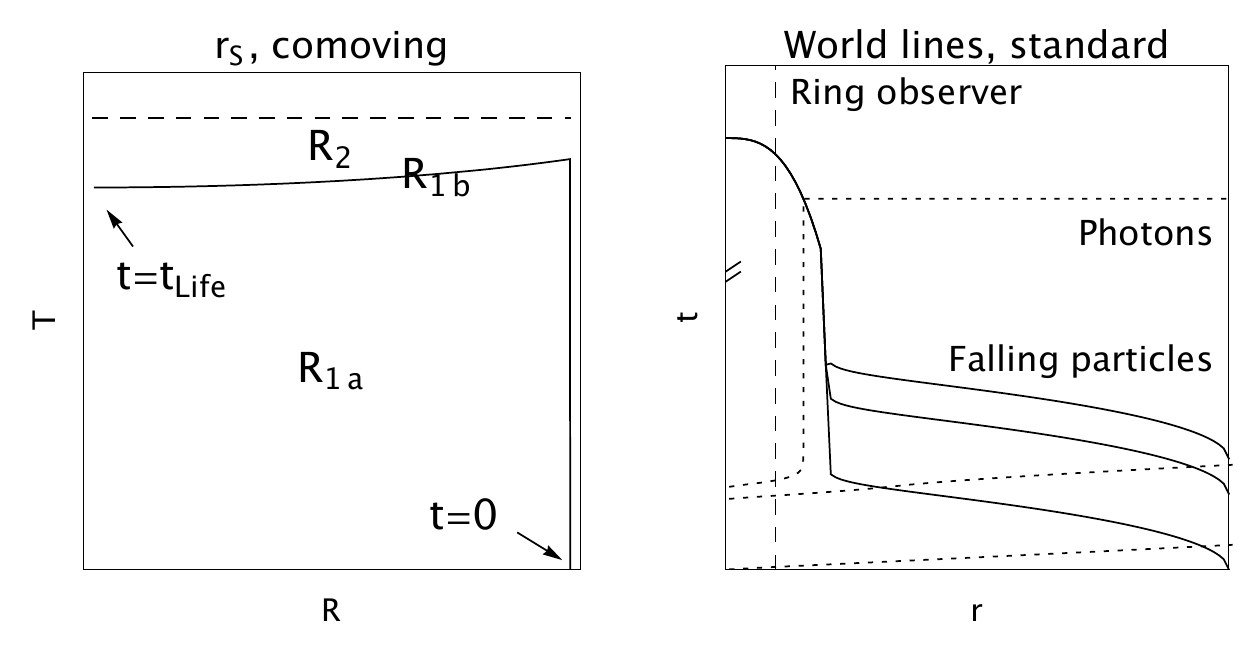}
\caption{\label{fig:7}
Left: the line $r_S(t)$ (solid) in comoving coordinates, dividing region $R_1$ into two subregions;
dashed line: location of singularity in the absence of evaporation.
Right: the geodesics of particles (solid) falling from $r=r_0$ at different times;
radially outgoing photons (dotted) starting at $r=0$ at different times;
the world line of an observer at constant $r$ (dashed).
The upper portion of the $t$ axis is compressed.
}
\end{figure}

{\bf Evaporation in comoving coordinates.}
We transform back to $(T,R)$ coordinates and plot $r_S(t)$ in the left panel of Figure~\ref{fig:7}.
The coordinates should now only be called ``comoving" for the region $R_{1a}$ interior to $r_S(t)$.
$R_{1a}$ is a subset of $R_1$ (see Figure~\ref{fig:1}(a)), and for the above evaporation model
the metric there is identical to that of classical collapse.
However, region $R_{1b}$ (a sliver too thin to discern) and all of $R_2$ do not apply to an evaporating black hole (in our model).

{\bf World lines, geodesics.}
The right panel of Figure~\ref{fig:7} shows various world lines.
The dashed world line is that of an observer on a ring of constant proper circumference $2\pi r$
(and not subject to destruction by the evaporation process).
The shape may look uninteresting.
The important point is the {\em existence} of a timelike world line at constant $r$,
starting inside the dustball and continuing outside it.
This existence follows from the fact that $A$ and $B$ remain positive, never changing signs,
in contrast to the (full) Schwarzschild metric, which does change signs at $r_S$.
The observer's wristwatch would asymptotically approach some value $T_o$ for much of $t<t_\t{Life}$,
then continue beyond $T_o$ as the ring emerges from the evaporating dustball, into a flattening spacetime.
In contrast, such a ring observer during classical collapse would experience a limited proper time
less than $T_\t{max}$ before encountering a singularity.

The figure also shows three infalling particles (solid lines).
The lowest line is for a particle on the dustball's original surface.
Two other particles fall from $r_0$ at later times: they almost catch up with the surface, but never enter the dustball,
since they trail behind the surface.

The figure also shows the null geodesics of three radially outgoing photons starting at $r=0$ at different times (dotted lines).
The geodesics all continue to infinity: none are trapped, no (absolute) horizon forms.
This also follows from the fact that $A$ and $B$ never change signs.

\begin{figure}[b]
\includegraphics[width=8.7cm,keepaspectratio]{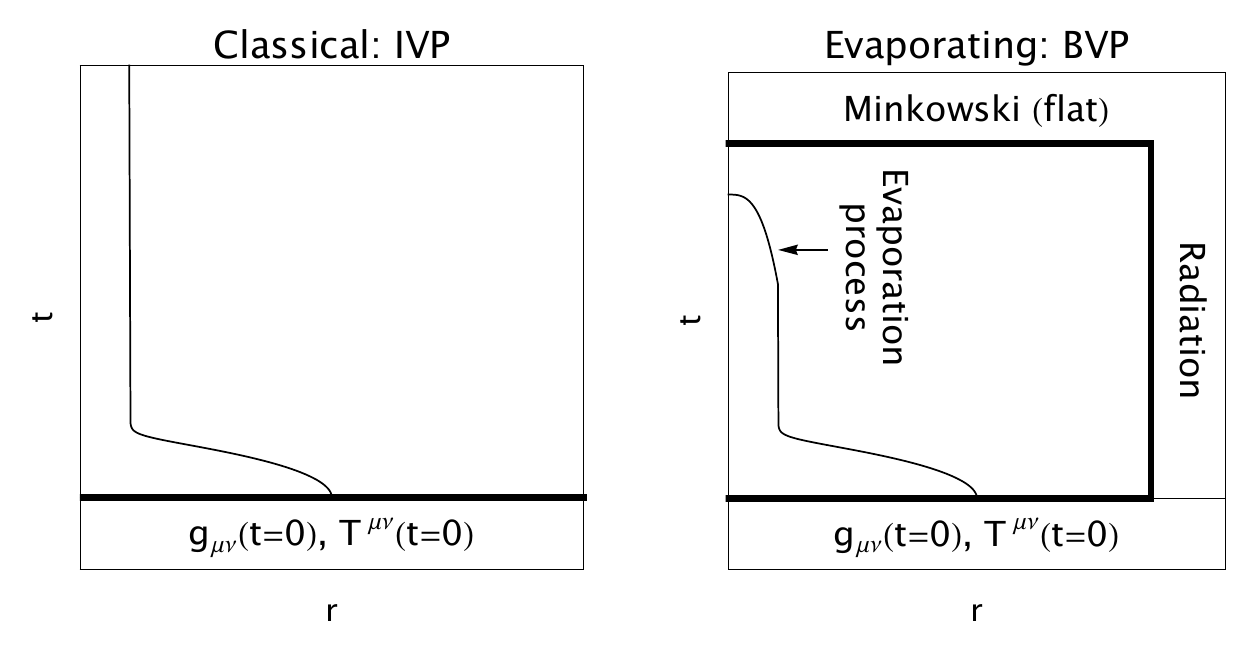}
\caption{\label{fig:8}
Left: classical collapse viewed as an initial-value problem.
Right: evaporating collapse viewed as a boundary-value problem.
Values and derivatives are specified on thick lines.
}
\end{figure}

\section{Discussion}

We started with Weinberg's implicitly defined metric,
derived explicit expressions for small and large $t$ and,
and generally characterized the metric.
Standard coordinates may be more intuitive when analyzing some problems,
such as the information paradox when bits approach the Schwarzschild radius of an evaporating black hole.

A simple evaporation model suggests that after the initial phase of collapse, nothing enters the region of the original dust,
and that neither singularity nor an (absolute) horizon forms.
Only a portion of the solution for classical collapse (region $R_{1a}$ in Figure~\ref{fig:7}) carries over to the evaporating case.
With the absence of region $R_{1b}$, the remaining $R_{1a}$ is disconnected from $R_2$, which is where the singularity resides for classical collapse.

Thus, we see no obvious reason to assume that a singularity forms.
With the absence of region $R_{1b}$, the burden of proof should rather be shifted onto anyone arguing that a singularity {\em does} form.
In general, when some coordinate system maps its finite $T$ to $t=\infty$, one must analyze which portion of a solution carries over to scenarios with finite $t$.

Hawking originally asserted collapse followed by evaporation \cite{Hawking-1975}.
Later studies of thin-shell collapse alongside evaporation saw the shell
``chasing its event horizon" and that an
``infalling flux of negative energy annihilates against the (still) collapsing matter,"
either generally or depending on the exact assumptions
\cite{Boulware-1976,Gerlach-1976,Alberghi-2001,Vachaspati-2007,Barcelo-2008,Dai-2016,Baccetti-2017}.
A dissenting paper argued that a shell does cross a horizon
\cite{Paranjape-2009}, and a recent review paper surveys a number of models \cite{Mann-2022}.

For a dustball, its sharp surface may well smear depending on the details of an evaporation model,
and so what exactly crosses the Schwarzschild radius is not well defined and may depend on
interpretation in terms of positive or negative energy fluxes.
However, within $t=1\,s$ the maximum distance a perturbation can propagate in finite time, $d_\t{max}$,
is vastly smaller than $r_S$ and so a negative energy flux would not continue on to $r=0$ but rather
``annihilate against the collapsing matter."

We elaborate on the difference between the two problems in Figure~\ref{fig:8}.
Classical collapse is here viewed as an initial-value problem,
with the metric and mass-energy distribution specified at $t=0$ (along with appropriate derivatives);
the solution is then obtained by evolving the field equations forward in time.
Evaporating collapse is viewed as a boundary-value problem on a spacetime region;
the boundary at $t=0$ is the same as for classical collapse, but at a later, post-evaporation time, the boundary is that of flat space;
the boundary at large $r$ is that of escaping Hawking radiation.
The boundary values then put constraints on the non-classical processes inside the boundary:
evaporation and the modified exterior Schwarzschild metric
(for example Vaidya null dust or an atmosphere with non-zero pressure).

\section{Appendix}

We build on Weinberg's treatment of classical dustball collapse, with some changes in notation and choice of parameters.
His comoving $(r,t)$ and standard $(r',t')$ become $(T,R)$ and $(t,r)$.
His $B$ and $A$ become $A$ and $B$.
His $R$ and $a$ become $Q$ and $r_0$.
Dustball parameters in Weinberg are $m$ and $a=r_0$, while this paper chooses $m$ and $n=r_0/(2m)$.
For convenience Weinberg introduces $k$, a rescaling of $\rho_0$, while this paper introduces $f$, a rescaling of $R$.

{\em Approximation of S(t)}.
In equation~\eqref{eq:tS2} we write $S=1/n+x$ and seek the limit as $x\rightarrow0$.
The first term diverges; it is rewritten as a logarithm, using a standard identity before taking the limit.
The limit of the second term is a constant.
The resulting expression of the form $t=a+b\ln(cx)$ is solved for $x$ to obtain $x(t)$ and $S(t)$.

{\em Approximation of $r_b(t)$}.
In equation~\eqref{eq:trb} we write $r_b=2m+x$ and proceed as for $S(t)$ above.

{\em Q as polynomial root.}
$Q$ is the solution of a quartic polynomial $Q^4+bQ^3+cQ^2+dQ+e$, where
$
b=-2,~
c=S(2-S) + (S-1)^2/n - r^2/(4m^2n^3),~
d=r^2/(2m^2n^3),~
e=-r^2/(4m^2n^3)
$
and the physically relevant root is a nested function of the coefficients $b,c,d$ and $e$:
$ Q=s-b/4-\sqrt{-(4s^2+2p-q/s)}/2$, where
$
p=c-3b^2/8,~
q=b^3/8-bc/2+d,~
\Delta_0=c^2-3bd+12e,~
\Delta_1=2c^3-9bcd+27b^2e+27d^2-72ce,~
Q=((\Delta_1+\sqrt{\Delta_1^2-4\Delta_0^3})/2)^{1/3},~
s=\sqrt{(Q+\Delta_0/Q)/3-2p/3}/2
$.

{\em Approximation $Q\approx k_2e^{-t/2m}+Q_f$.}
For large $t$, we write $S=1/n+x$, with $x=k_0e^{-t/2m}$ from the approximation of $S(t)$.
Only the coefficient $c$ depends on $x$; we write it as $c=c_1+c_2x,$ discarding the $x^2$ term.
We let this linear approximation of $c$ propagate up the nested functions that result in $Q$,
discarding terms higher than linear in $x$, and using approximations like
$(a_1+a_2x)^{1/n}  \approx  a_1^{1/n}+a_2/(n\,a_1^{(n-1)/n})$.
This is conceptually straightforward, but tedious,
and we obtain $k_2=s_2-m_2/(4\sqrt{m_1})$, where
$
m_1=-2p_1-q_1/s_1-4s_1^2,~
m_2=-2p_2-q_2/s_1+q_1s_2/s_1^2-8s_1s_2,~
s_1=\sqrt{j_1}/2,~
s_2=j_2/(\sqrt{j_1}/4),~
j_1=(-2p_1+Q_1+\Delta_{01}/Q_1)/3,~
j_2=(-2p_2+Q_2-Q_2\Delta_{01}/Q_1^2+\Delta_{02}/Q_1)/3,~
Q_1=h_1^{1/3},~
Q_2=h_2/(3h_1^{2/3}),~
h_1=(\Delta_{11}+g_1)/2,~
h_2=(\Delta_{12}+g_2)/2,~
g_1=\sqrt{f_1},~
g_2=f_2/(2\sqrt{f_1}),~
\Delta_{01}=c_1^2-3bd+12e,~
\Delta_{02}=2c_1c_2,~
\Delta_{11}=2c_1^3-9bc_1d+27b^23+27d^2-72c_1e,~
\Delta_{12}=6c_1^2c_2-9bc_2d-72c_2e,~
p_1=c_1-3b^2/8,~
p_2=c_2,~
q_1=b^3/8-bc_1/2+d,~
q_2=-bc_2/2
$.

{\em Approximation of $B(t)$ for $r<r_S$.}
In equation~\eqref{eq:B} we use approximation of $Q$ for $Q$ and in $f=r/(2nmQ_f)$,
then simplify, keeping linear terms, as when deriving the approximation of $Q$.

{\em Approximation of $A(t)$ and $c_r(t)$ for $r<r_S.$}
In equations~\eqref{eq:A} and \eqref{eq:cr} we use the already derived exponential approximation for the factor $nS-1$
and final values for $Q,S$ and $f$.

\bibliographystyle{apsrev4-1}
\bibliography{bernander}
\end{document}